\documentclass[manuscript]{aastex}
\usepackage{amssymb}

\usepackage{amsmath}

\slugcomment{Not to appear in Nonlearned J., 45.}

\shorttitle{Fine Magnetic Characteristics of a Light Bridge}
\shortauthors{S. Liu et al.}

\begin{document}


\title{Fine Magnetic Characteristics of a Light Bridge Observed by Hinode}

\author{S. Liu\altaffilmark{1,2}, D. Liu\altaffilmark{1}}

\affil{
$^{1}$Liao Ning University,
        Shenyang, China\\
$^{2}$National Astronomical Observatory, \\Chinese Academy of
Sciences,
        Beijing, China}

\email{liud@lnu.edu.cn}
\email{lius@nao.cas.cn}

\altaffiltext{1}{key Laboratory of Solar Activity}


\begin{abstract}
Light bridge (LB) is bright structure crossing the umbra of
sunspots and associated to the breakup or assembly of sunspots. In
this paper, a LB is presented and studied using the observatory data
obtained by {\it Hinode} satellites. Force-free factor ($\alpha$)
and the z-component of current ($J_{z}$) and tension force ($T_{z}$)
are calculated basing on the vector magnetograms observed by
Spectro-Polarimeter (SP) of the Solar Optical Telescope (SOT) on
board {\it Hinode}. It is found that the amplitudes of $\alpha$ and
$J_{z}$ of LB are generally larger than those of umbra. It is found
that there are two signs of $J_{z}$ along LB, which are divided at
near the middle position of LB. It is found that the amplitudes of
$T_{z}$ of LB are smaller than those of umbra and there are changes
of sign of $T_{z}$ between the boundary of LB and umbra.
\textbf{Through comparisons and investigations, it suggest that LB and umbra maybe two different magnetic systems,
which is a necessary condition for interaction magnetic reconnection.}
\end{abstract}


\section{Introduction}

Sunspots dominated by strong magnetic field with the amplitude about
K-Gauss are main features of Sun. There are some magnetic structure
in Sunspots, such as umbra, penumbra, filamentary structure, umbral
dots (UCs). Light bridges (LBs), which are bright, long, and narrow
feature penetrating or crossing the umbra during the evolution of
sunspots, are also one of the fundamental magnetic structures in
sunspots. LBs are associated to the breakup of sunspots in the decay
or the assembly of sunspots in complex active regions \citep{bra64,
vas73, gar87}. LBs can be classified LB as
"photospheric,""penumbral," and "umbral" LB according its intensity
and fine structure \citep{mul79}. According to their width,
\citet{sob93, sob94} classified LBs as strong LB, which separate
umbral core and is further distinguished as photospheric or
penumbral, and faint LB, which is faint narrow lane with in the
umbra and most likely consists of umbral dots.
Recently, the high resolution observation revealed there are more fine structures in LB,
such as dark central lanes running along the length of LB, bright grains along length of LB, narrow dark lanes
that can separated LB oriented perpendicular to the length of LB \citep{ber03, shi09}.

The previous studies show that the structures of LB are evident different from
those of umbra \citep{lek97, jur06, spr06}.
Magnetic field in LB is revealed weaker and more inclined than than
in the neighboring umbra \citep{rue95, lek97, jur06}. Moreover, by a
detail analysis of the Stokes spectra \citep{jur06}, it is found
that the field strengths and inclinations increase and decrease with
height, which may suggest a canopy-like structure above the LB. At
present, the formation and magnetic properties of LB are not known
completely. A physical mechanism to explain the formation of LB is
that field-free convection penetrates umbra from sub-photosphere and
forms a cusp-like magnetic field \citep{spr06}. \citet{kat07b}
revealed the formation of a LB due to the intrusion of umbral dots
basing on data obtained from H$inode$ satellite. Based on H$inode$
observation of the magnetic field in a LB accompanied by
long-lasting chromospheric plasm ejections, \citet{shi09} suggest
that current-carrying highly twisted magnetic flux tubes are trapped
below a cusp-shape magnetic structure along the LB. The universal
solar activities related to LB are remarkable plasma ejections or
H$\alpha$ surge in chromosphere along LB \citep{roy73, asa01, bha07,
shi09}. The bright enhancement over the site of LB in 1600 \AA
~images and heating of coronal loops in 171 \AA ~images from
Transition Region and Coronal Explorer (TRACE) was founded
recently \citep{ber03, kat07a}, which may suggest that LB is a
steady heat source in the chromosphere. There are also corona
activities may related to LB. For example, \citet{liu11} reported a
coronal jet that may related to the interaction between LB and
umbra.

The equilibrium structures of sunspots are dominated
by magnetic forces, since there are low-$\beta$ plasmas in the most part of sunspots.
The formation and disappearance of sunspot magnetic field are one of the key
problems of solar physics. Because of magnetic freezing phenomenon
sunspot magnetic field can not disappear through magnetic diffusion.
The fine magnetic structures and features of sunspot become
an important and essential aspect to study the formation and
disappearance of sunspot magnetic field. LB is one of the
fundamental and obvious magnetic structures in sunspots, hence the knowledge of
magnetic properties of LB is an important channel to study sunspot
magnetic field. The previous studies have reported some basic
information about LB magnetic field, however the high spatial
resolution vector magnetic field observed by SP/SOT on board {\it
Hinode} give us an unprecedented opportunity to reveal LB magnetic
field.

In this paper, some basic physical quantities, which related
to magnetic field such as $\alpha$, $J_{z}$ and $T_{z}$ are
studied and investigated.
$\alpha$ ($\alpha(\emph{r})=(\nabla \times \emph{B})/\emph{B}$) indicate the strength and direction of
twist of local magnetic field lines \citep{tiw09, su10, zhang10}. The extent of twist can affect the stabilities of
magnetic field lines, such as pinch instability \citep{ryu08}. $J_{z}$ demonstrate the strength and direction of
current, which indirectly related to the topology of magnetic field lines \citep{wang08, zhang10, rav11}.
For exmaple, one mechanism proposed to explain vertical electric currents at the photosphere is that
the surface flows dray magnetic field lines into non-potential configurations if the field are "frozen"
to the plasmas \citep{tan73, sch94}, thus the currents indirectly indicate the distributions or redistributions of magnetic field.
There are various forces (such as gravity, gas pressure, Lorentz force)
that dominate the equilibrium of sunspot plasmas, as there are strange magnetic field in the sunspot the forces associated to magnetic field
should play an especial important role in keeping the equilibrium of sunspot plasmas. The sunspot is usually modeled as a magnetic flux rope
where the outer photospheric plasma pressure balances the magnetic and plasma pressure inside the flux rope.
$T_{z}$ is the force related to the strength and direction of bent magnetic field lines. The equilibrium of sunspot can
become unstable if the radius of curvature of field line is shorter than a certain value \citep{ven93}. $T_{z}$ may also create the changes of
magnetic topology if there are some instabilities among the corresponding magnetic and thermal circumstances
of plasmas \citep{ven93, ven10}. The above parameters are magnetogram dependent intensely, some more information contained in
magnetograms can be revealed through these magnetic parameters. Such as current distribution and the properties of twist can indirectly manifest the topology of
magnetic field.

The paper is organized as follows: firstly, the description of
observations and data used will be introduced in
Section~\ref{S-Obser and Data}; secondly, the results will be shown
in Section~\ref{S-Results}; at last, the short discussions and
conclusions will be given in section~\ref{S-Conl}.

\section{Observations and Data Reduction}
\label{S-Obser and Data}

LB studied here belongs to the lead negative sunspot of NOAA 11271,
which is a $\beta$/$\beta$$\gamma$$\delta$ active region. LBs are
observed during the time from 08:05:05 to 10:05:06 UT on 19 Aug
2011, when the active region locates about N16E26 in heliographic
coordinates. The observatory data used to study this LB were
obtained by Solar Optical Telescope (SOT) on board H$inode$
\citep{kos07, tsu08}. G-band and Ca II H with spatial resolution of
0.1 arcsec and vector magnetograms with spatial resolution 0.16
arcsec obtained by SOT/H$inode$ are used in this work. Where G-band
and Ca II are observed by Broadband filter of SOT and vector
magnetograms are observed by Spectro-Polarimeter (SP) of SOT. For
G-band and Ca II data, the data processing in the work are all based
on standard solar software (SSW {\it e.g}, fg\_prep.pro). For
example, dark subtraction, flat fielding, the correction of bad
pixels and cosmic-ray removal were done for filtergram images
obtained by SOT.
The parameters relevant to the
vector magnetic field, which are derived from the inversion of the full Stokes profiles based
on the assumption of the Milne-Eddington (ME) atmospheric model.
For the vector magnetogram, the data are load down
from the web of
http://bdm.iszf.irk.ru/sfq$\_$hinode/SFQ$\_$Hinode.htm. These data
include $B_x$,$B_y$ and $B_z$ as output, where the azimuth ambiguity
of the transverse field were deal with Super Fast and Quality (SFQ)
method and the projection effect are considered in
this SFQ method \citep{rud14}.
\section{Results}
\label{S-Results}

\begin{figure}
   \centerline{\includegraphics[width=1.0\textwidth,clip=]{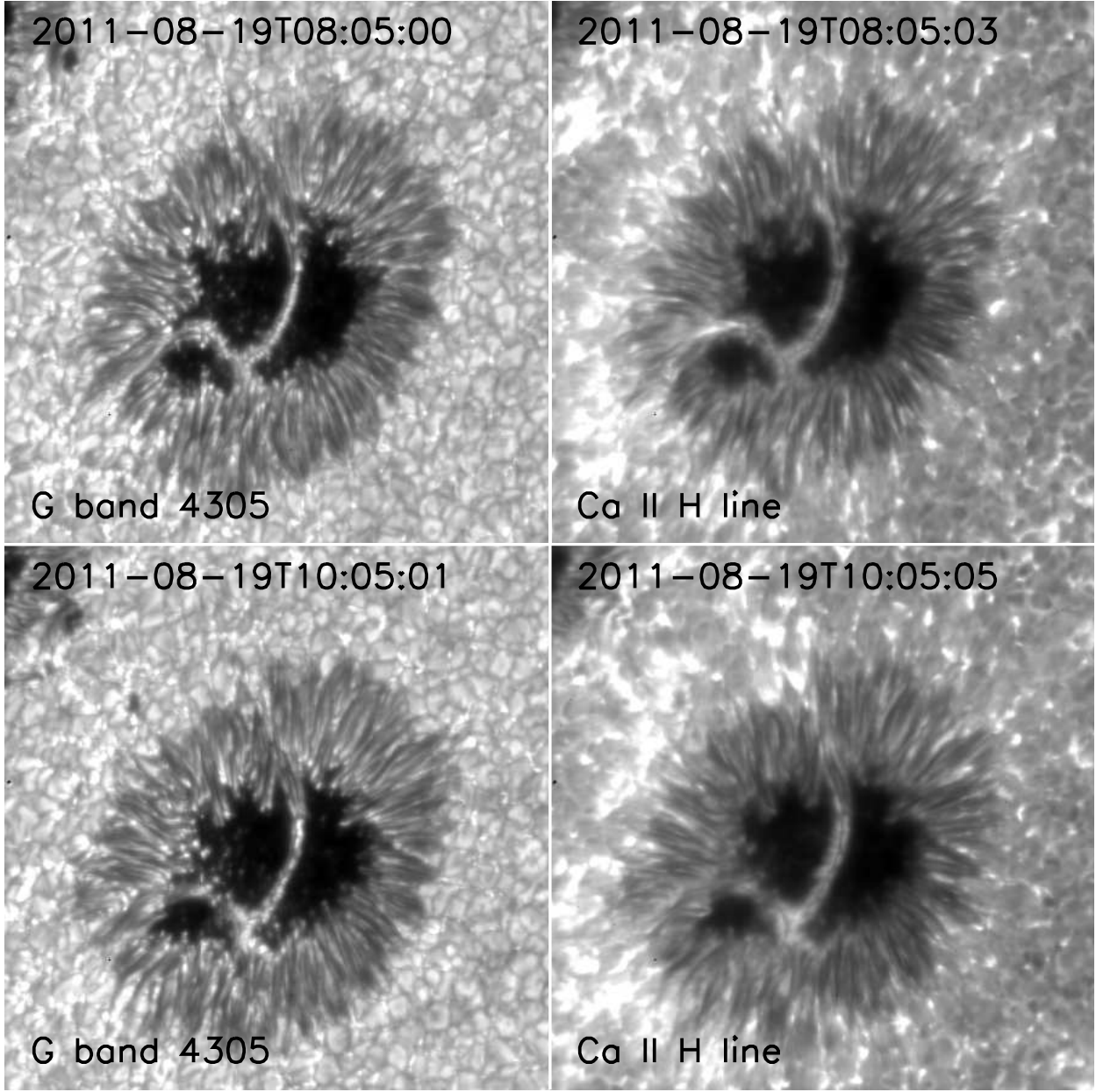}}

   \caption{Two images in the first column are G band 4305 observed at 08:05:00 and 10:05:01 UT on 18 Aug 2011, respectively.
Two images in the second column are Ca II observed at 08:05:00 and
10:05:01 UT on 18 Aug 2011, respectively. } \label{Fig1}
\end{figure}

\begin{figure}
   \centerline{\includegraphics[width=1.0\textwidth,clip=]{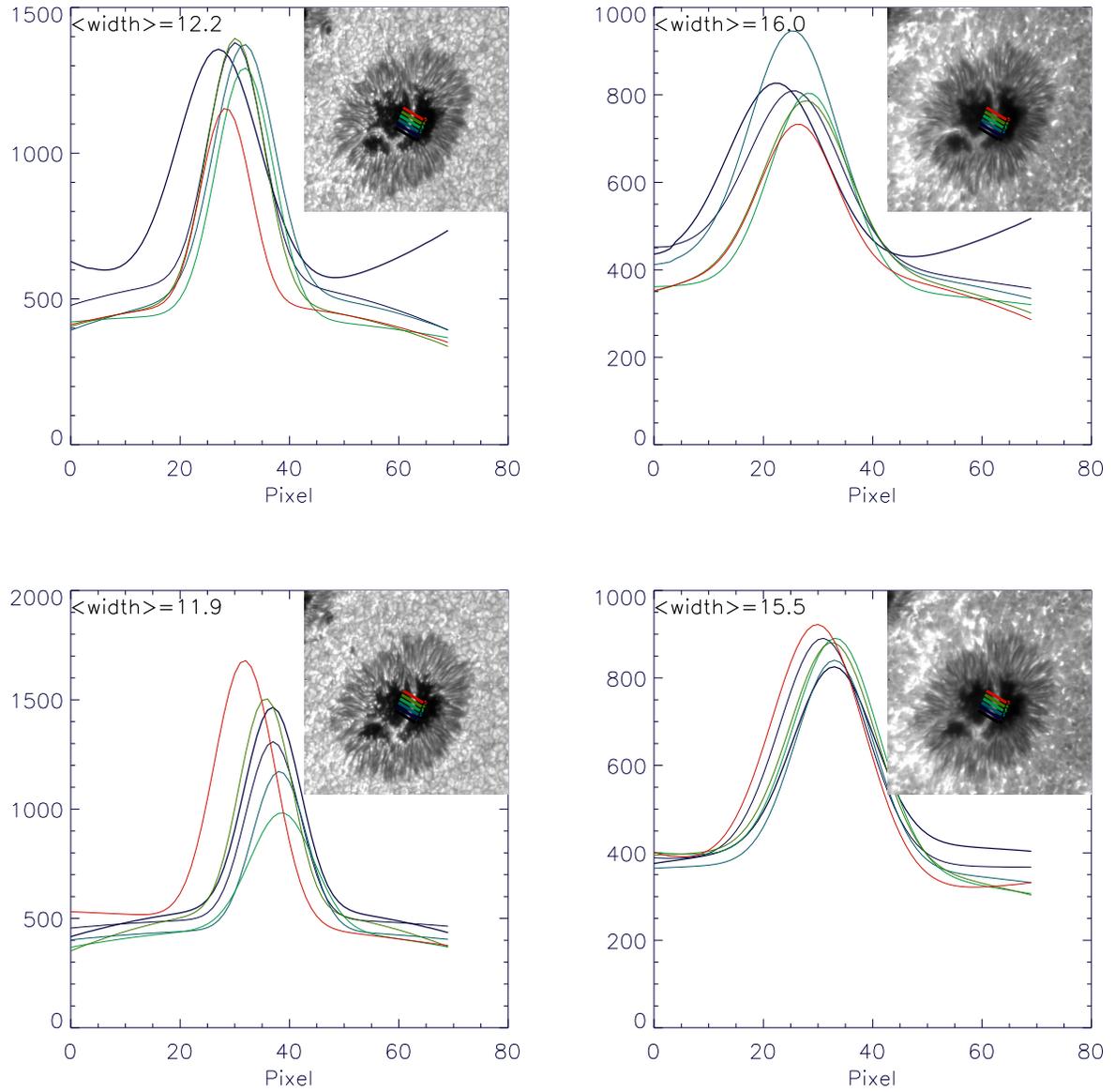}}

   \caption{The figures to show the width of LB in images of G band 4305 and Ca II. The gray-scales in each column correspond
   images of \ref{Fig1}, in each image there are six lines crossing LB that are plotted to display the width of LB, here the
   average of width with pixel unit are labeled. } \label{Fig1-add}
\end{figure}

Fig~\ref{Fig1} shows G band 4305 and Ca II H images of this LB
observed at two different time of 08:05:00 and 10:05:01 UT on 18 Aug
2011, respectively. Where the field of view is 45 arcsec $\times$ 45
arcsec, the first column and the second column shows G band 4305
images and Ca II H images, respectively. From this Figure, it can be
seen that this LB should be classified "umbral" LB according its
intensity and fine structure, also it should be regarded as a strong
one, since this LB penetrates the umbra completely and separates
umbral core into two parts evidently. When it is seen more
carefully, a evident ridge structure in the middle of LB along its
length direction can be seen. This ridge structure display the dark
features in these images, however it is more evidently displayed in
Ca II H images. This maybe because the ridge structure locates at
nearly the center position of LB in Ca II H images, while in G band
images the position where ridge locates departed from the center
position of LB. The ridge structure in G band images close to the
west boundary of LB and the corresponding umbra. This observation
confirm once again the previous results that the dark central lanes
running along the length of LB.The width of LBs, which are $\sim$
900/1150 km for Ca II H/G band observations, are calculated from these images Fig \ref{Fig1-add}.
In Fig \ref{Fig1-add}, six lines crossing individual LB corresponding Fig \ref{Fig1} images
are selected to calculated the width of LB, through calculation using  GAUSS-FIT and regarding halfwidth as the width of LB,
the color fit lines correspond individual line with the same color, respectively.
The average of widths of LB in Ca II H/G band images about 12.1/15.8 pixel with the pixel size of 0.1 arcsec.

\begin{figure}
   \centerline{\includegraphics[width=1.\textwidth,clip=]{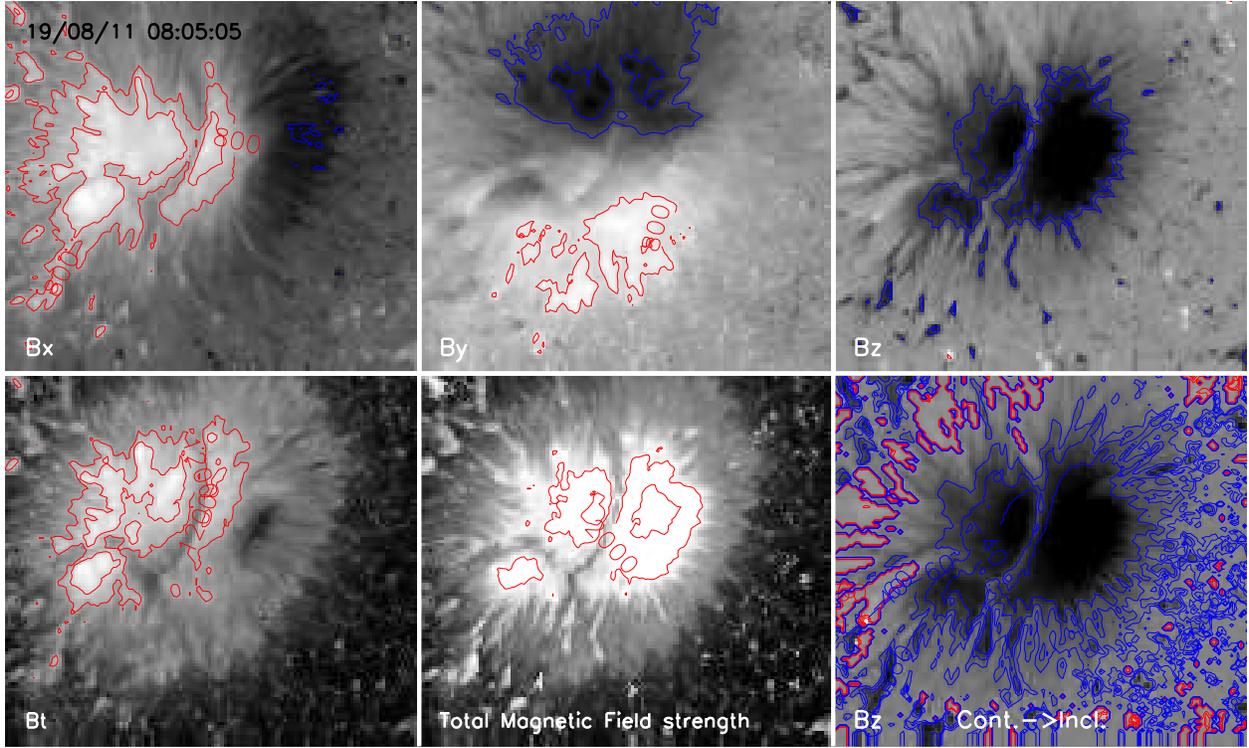}}

   \caption{The images of magnetic component of $B_{x}$ with contour levels $\pm$800 and 1200 G, $B_{y}$ with contour levels $\pm$800 and 1200 G,
    $B_{z}$ with contour levels $\pm$1200 and 1500 G, $B_{t}$ (transverse magnetic field) with contour levels $\pm$1200 and 1500 G,
    $B_{tot.}$ (total magnetic field strength) with contour levels $\pm$2000 and 2500 G and
   the contour of inclination angle plotted on $B_{z}$ map with the contours levels $\pm$40$^{\circ}$, 50$^{\circ}$ and red/blue
contours represent positive/negative values of inclination angle, the corresponding observation time is label. }
\label{Fig2}
\end{figure}

In order to investigate LB magnetic field, Fig~\ref{Fig2} shows
magnetic components of $B_{x}$, $B_{y}$, $B_{z}$, the transverse magnetic field $B_{t}$ and
total magnetic field strength $B_{tot.}$ for this LB observed
time 08:05:05 UT. Basing on magnetic components the inclination angles of this
LB are calculated, and the distribution of inclination angles
(namely, atan($B_{t}/B_{z}$), where transverse magnetic field
$B_{t}$ = $\sqrt{ B_{x}^{2}+B_{y}^{2}}$) indicated by contour lines
are plotted on the grey-scale map of $B_{z}$ in the last column of
Fig~\ref{Fig2}. From Fig~\ref{Fig2} it is can be found that the LB
structures are seen clearly in image of each magnetic components of
$B_{x}$, $B_{y}$ and $B_{z}$ and each deduced magnetic component.
Here, the magnetic component of $B_{x}$ with contour levels $\pm$800 and 1200 G, $B_{y}$ with contour levels $\pm$800 and 1200 G,
$B_{z}$ with contour levels $\pm$1200 and 1500 G, $B_{t}$ (transverse magnetic field)  with contour levels $\pm$1200 and 1500 G,
$B$ (total magnetic field strength) with contour levels $\pm$2000 and 2500 G, respectively.
But for the inclination angles, the amplitudes of magnetic components
($B_{x}$, $B_{y}$, $B_{z}$, $B_{t}$ and $B_{tot.}$) of LB are all smaller than those of its neighboring
umbra. As for the inclination angles, it is found that the
inclination angles of LB are larger than those of neighboring umbra.
Hence, the large value of inclination angles are shown as contour
lines in Fig~\ref{Fig2}, where the contours are $\pm$40$^{\circ}$,
50$^{\circ}$ and red/blue contours represent positive/negative
values of inclination angle, respectively. From the distribution of
inclination angles, it can be found that the amplitude of
inclination angles of LB are comparable to those of penumbra and a
part of quiet regions. While there are less contour lines, which
means less large value of inclination angles, exist at the region of
umbra. In the images of $B_{x}$, $B_{y}$, $B_{z}$, $B_{t}$ and $B_{tot.}$,
the relation between longitudinal and transverse
magnetic field can not be clearly seen, but ratio of transverse
field to longitudinal field that contained indirectly in inclination
angles is higher than those of umbra, which also means that the
magnetic field of LB are more inclined than those of umbra. However,
here it should be noted that the total intensities of magnetic field
of LB are weaker than those of umbra undoubtedly (see from the distributions of
$B_{tot.}$).

To study magnetic field of LB, physical quantity force-free factor
($\alpha$) and z-component of current ($J_{z}$) and tension force
($T_{z}$) are calculated basing on the vector magnetograms observed
at three different time. The definitions of $\alpha$, $J_{z}$ and
$T_{z}$ are as follows:

From the basic electromagnetic equations force free factor
($\alpha$), current ($\emph{\textbf{J}}$) and Lorentz force
($\emph{\textbf{F}}$) can be expressed:
\begin{equation}
\label{alpha}\alpha \emph{\textbf{B}} = \bigtriangledown \times
\emph{\textbf{B}} = (\dfrac{\partial B_{z}}{\partial
y}-\dfrac{\partial B_{y}}{\partial
z})\emph{\textbf{i}}+(\dfrac{\partial B_{x}}{\partial
z}-\dfrac{\partial B_{z}}{\partial
x})\emph{\textbf{j}}+(\dfrac{\partial B_{y}}{\partial
x}-\dfrac{\partial B_{x}}{\partial y})\emph{\textbf{k}},
\end{equation}

\begin{equation}
\label{J} \emph{\textbf{J}}=\dfrac{1}{\mu_{0}}\bigtriangledown
\times \emph{\textbf{B}}=\dfrac{1}{\mu_{0}}[(\dfrac{\partial
B_{z}}{\partial y}-\dfrac{\partial B_{y}}{\partial
z})\textbf{\emph{i}}+(\dfrac{\partial B_{x}}{\partial
z}-\dfrac{\partial B_{z}}{\partial
x})\textbf{\emph{j}}+(\dfrac{\partial B_{y}}{\partial
x}-\dfrac{\partial B_{x}}{\partial y})\textbf{\emph{k}}],
\end{equation}

\begin{equation}
\label{F} \emph{\textbf{F}}=\dfrac{(\emph{\textbf{B}}\centerdot
\bigtriangledown )\emph{\textbf{B}}}{\mu_{0}}-\dfrac{\bigtriangledown
(\emph{\textbf{B}}\centerdot \emph{\textbf{B}})}{2\mu_{0}}).
\end{equation}
Where $\alpha$ can indicate the twist and current of field line at
some extent. $\emph{\textbf{F}}$ can demonstrate the equilibrium of
sunspots structure \citep{ven10} at some extent. In
equation~\ref{F}, the first term is the tension force
($\emph{\textbf{T}}$) and the second term represents the force due
to magnetic pressure. Hence $\alpha$, z-component of current
($J_{z}$) and tension force ($T_{z}$) can be expressed as follows:

\begin{equation}
\label{alpha1} \alpha = (\dfrac{\partial B_{y}}{\partial
x}-\dfrac{\partial B_{x}}{\partial y})/B_{z},
\end{equation}
\begin{equation}
\label{Jz} J_{z}=\dfrac{1}{\mu_{0}}(\dfrac{\partial B_{y}}{\partial x}-\dfrac{\partial
B_{x}}{\partial y}).
\end{equation}
\begin{equation}
\label{Tz} T_{z}=\dfrac{1}{\mu_{0}}[B_{x}\dfrac{\partial
B_{z}}{\partial x}+B_{y}\dfrac{\partial B_{z}}{\partial
y}-B_{z}(\dfrac{B_{x}}{\partial x}+\dfrac{ B_{y}}{\partial y})].
\end{equation}
At last, $\alpha$, $J_{z}$ and $T_{z}$ all can be obtained on the
photosphere, since the horizontal derivatives of the vector magnetic
field observed on the photosphere can be calculated.

\begin{figure}
   \centerline{\includegraphics[width=1.\textwidth,clip=]{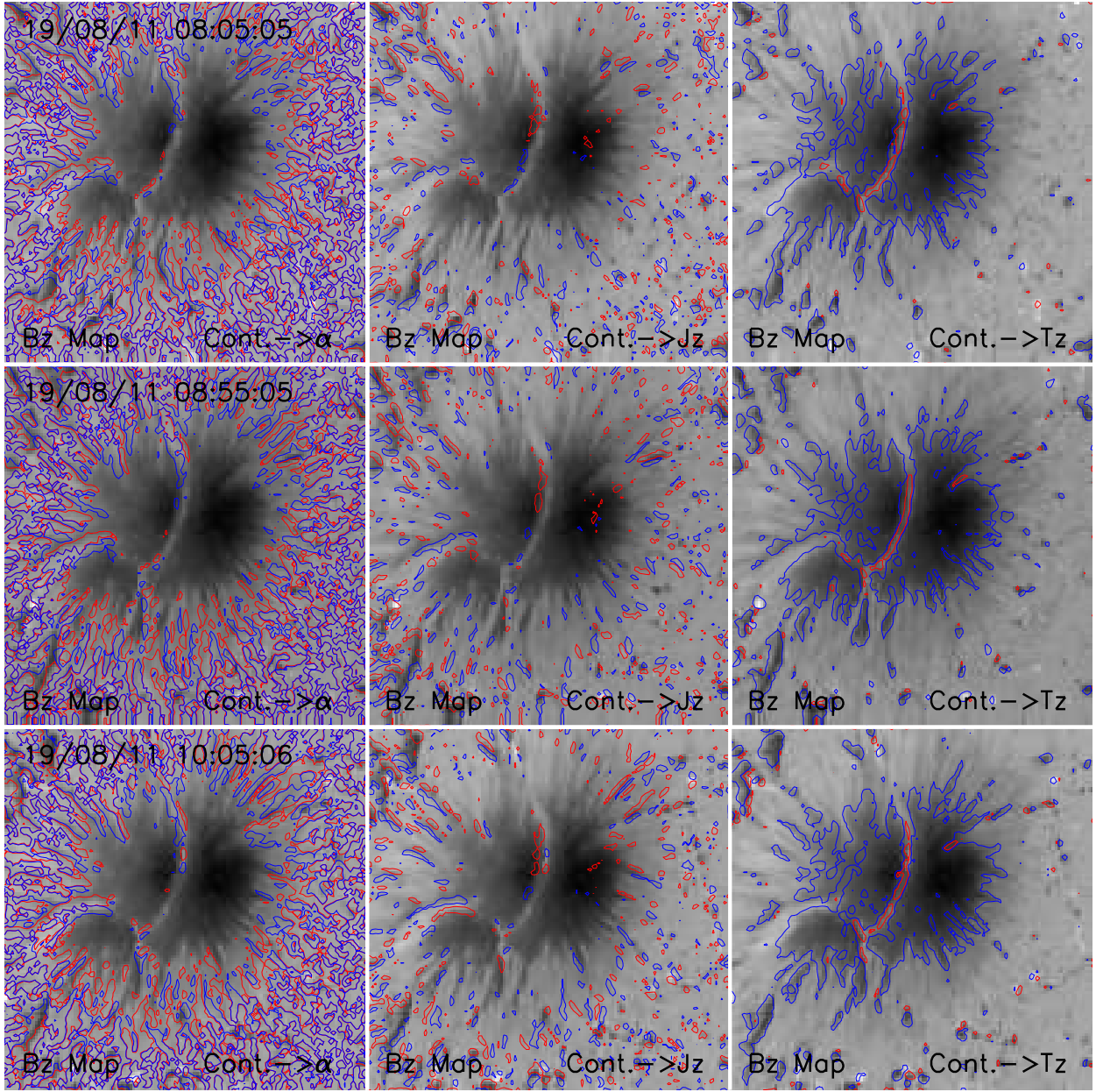}}

   \caption{The background image are $B_{z}$ map and  contour lines
represent for the distribution of $\alpha$, $J_{z}$ and $T_{z}$,
respectively. For $\alpha$ in the first column, the contours are
$\pm$0.008, 0.009 $m^{-1}$ and red/blue contours represent
positive/negative values. For $J_{z}$ in the second column, the
contours are $\pm$0.08, 0.09 $Am^{-2}$ and red/blue contours
represent positive/negative values. For $T_{z}$ in the third column,
the contours are $\pm$0.11, 0.12 $G^{2}m^{-1}$ and red/blue contours
represent positive/negative values. For vector $J_{z}$ and $T_{z}$,
the positive/negative values means the direction of this vector is
up/down to the photosphere, respectively} \label{Fig3}
\end{figure}

\begin{figure}
   \centerline{\includegraphics[width=1.\textwidth,clip=]{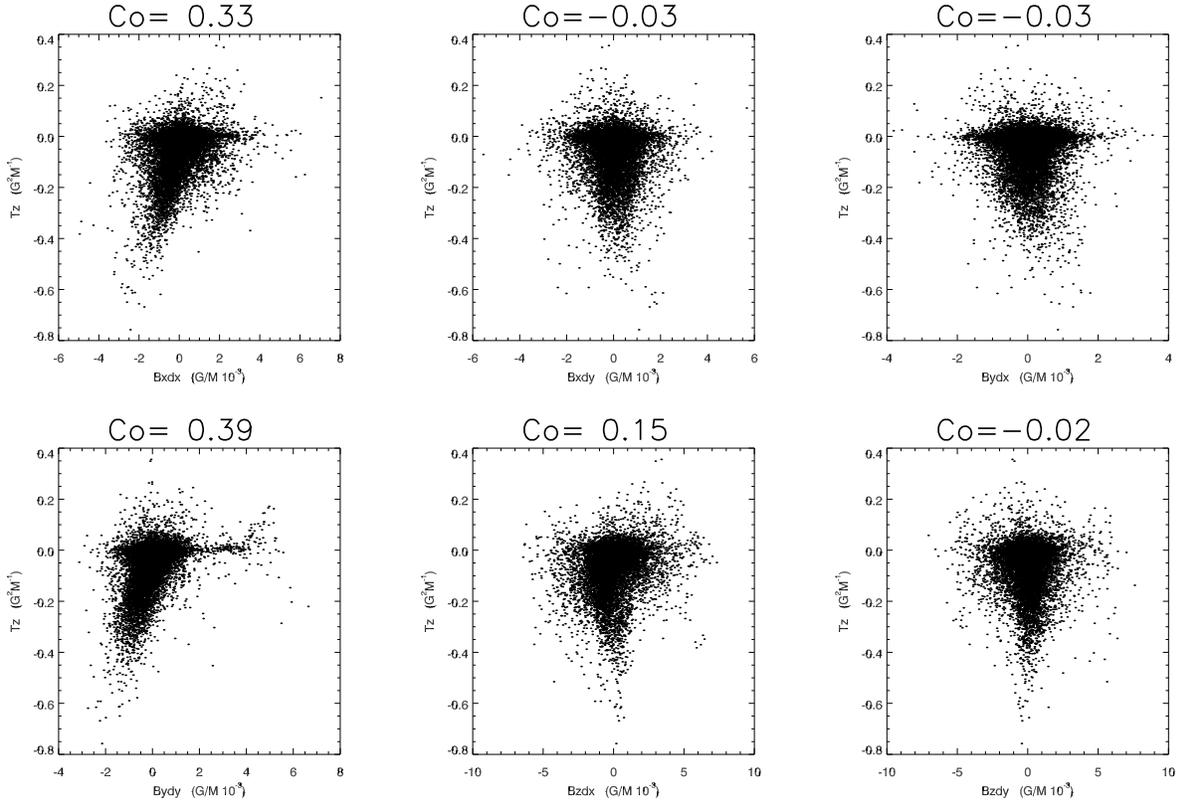}}

   \caption{The scatter diagrams to show the relationships between $T_{z}$ and the gradient of magnetic
   components, where the gradient of magnetic components are labeled correspondingly. } \label{Fig3-add}
\end{figure}

\begin{figure}
   \centerline{\includegraphics[width=1.\textwidth,clip=]{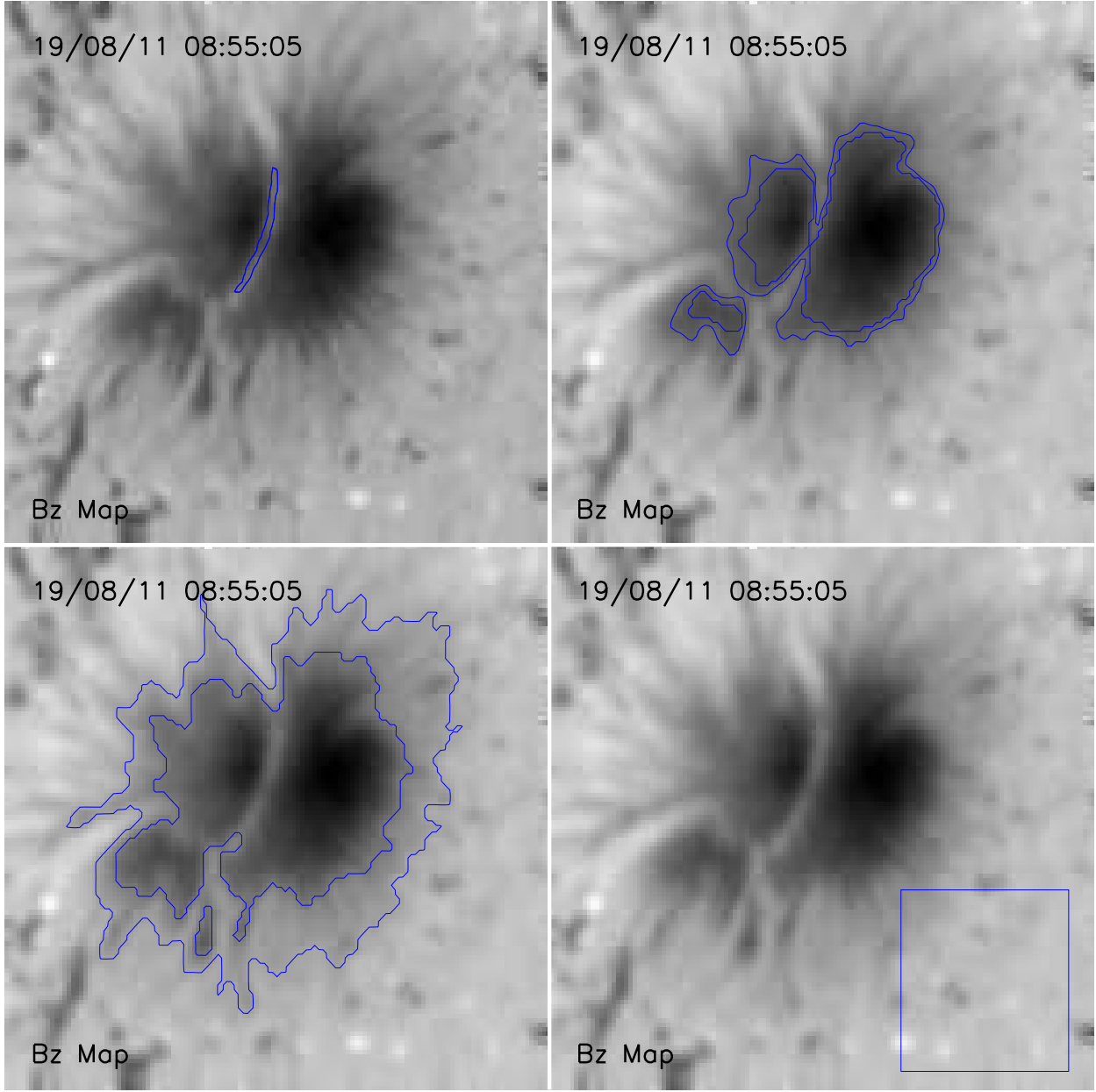}}

   \caption{It shows the selected region of LB, umbra, penumbra and quiet region labeled by blue closed lines, here
the  semi artificial method (IDL region-grow) are used to choose interested sub-region. } \label{alp-jz-tz-exma}
\end{figure}

Fig~\ref{Fig3} shows the images of $\alpha$, $J_{z}$ and $T_{z}$
calculated from vector magnetic field at three different observed
time. For $\alpha$ in the first column, the contours are $\pm$0.008,
0.009 $m^{-1}$ and red/blue contours indicate positive/negative
values of $\alpha$. It is found that $\alpha$ of LB are basic larger
than those of its neighboring penumbra and umbra, however a large
mount of large values of $\alpha$ are also dispersed in quiet
regions. For $J_{z}$ second column, the contours are $\pm$0.08, 0.09
$Am^{-2}$ and red/blue contours represent positive/negative values,
where the the positive/negative values means the direction of
$J_{z}$ is up/down to the photosphere, respectively. On the whole it
is found that $J_{z}$ of LB are larger than those of other parts
including penumbra and umbra. Generally, it is found that one part
of $J_{z}$ of LB have positive values and the other part have
negative values, and this two are are divided at about the middle of
LB along its length roughly (especially, seen evidently in the up
panel in the second colume). For $T_{z}$ third column, the contours
are $\pm$0.11, 0.12 $G^{2}m^{-1}$ and red/blue contours also
represent positive/negative values of $\alpha$, and
positive/negative contour lines also mean the direction of $T_{z}$
is up/down to the photosphere, respectively. Here the contour lines
of $T_{z}$, which surround the boundary of LB and umbra, are plotted
in this figure. It can be found that the direction of $T_{z}$ is
opposite in LB and umbra. Additionally, an interested thing is that
the change of direction of $T_{z}$ is exactly at the boundary
between LB and umbra. This demonstrates that Lorentz force of
boundary of LB has the trend up to the photosphere, while Lorentz
force of boundary of umbra has the trend down to the photosphere.
Hence, from the distribution of $T_{z}$ it can be found that
magnetic system of LB and umbra should be regarded as two different
systems.
From Lorentz force equation (Eq.3), it can be seen that $T_{z}$
possibly is sensitive to the gradient of magnetic components.
So in Fig \ref{Fig3-add} the relationships between
$T_{z}$ and the gradient of magnetic components are shown by scatter diagrams tentatively, and the correlation coefficients are labeled respectively.
It can be found that
the large amplitudes $T_{z}$ (most of parts are negative values) correspond to small amplitudes of magnetic component
gradient on the whole. While for small amplitudes of $T_{z}$ there are
various amplitudes of magnetic component gradient can appear,
then the amplitudes of magnetic components ($B_{x}$,$B_{y}$ and $B_{z}$ in Eq.3)
may play relatively important contributions to $T_{z}$.
The amplitude of $\alpha$, $J_{z}$ and $T_{z}$ in LB and
other parts are calculated to see their differences related to
position. Fig \ref{alp-jz-tz-exma} shows the selected region of LB, umbra, penumbra and quiet region that labeled by blue closed lines, here
the  semi artificial method (IDL region-grow) are used to choose interested sub-region.
It is found that the amplitude of $\alpha$ is about 0.007,
0.002, 0.004 and 0.100 $m^{-1}$ in LB, umbra, penumbra and quiet
region, respectively. It is found that the amplitude of $J_{z}$ is
about 0.05, 0.03, 0.04 and 0.05 $Am^{-2}$ in LB, umbra, penumbra and
quiet region, respectively. It is found that the amplitude of
$T_{z}$ is about 0.12, 0.30, 0.10 and 0.01 $G^{2}m^{-1}$ in LB,
umbra, penumbra and quiet region, respectively.

\section{Discussions and Conclusions }
\label{S-Conl}

LBs are common magnetic feature during the evolution of sunspot,
hence the magnetic properties of LBs can contribute the knowledge of
sunspot magnetic field. Additionally, there are exist plentiful magnetic
activities along LB and at the boundaries between LB and its
neighboring umbra, such as chromospheric brighten, H$\alpha$ surge,
corona jet and so on. Because the widths of LB are too narrow to be studied accurately
by all current observations all most. Hence
the formation, disappearances and properties of LB and also the
interactions between LB and umbra are not known adequately. At
present, the observatory of {\it Hinode} satellites is a high
spatial resolution one. Although the spatial resolution of {\it
Hinode} data maybe not high enough to study LB accurately, it
gives us a good chance to maximum disclosure the fine properties of magnetic field and its related parameters.

In this paper, a LB existing in the lead sunspot of NOAA 11271
observed by {\it Hinode} satellites are presented and studied. Due
to observations limitations (here 3 hours observation and gives 3 magnetograms available), the evolutions of LB are not
reported in this paper, hence the object of this study should be
regarded as a static LB neglecting its evolutions. The fine magnetic
structures and features about LB are the main objects that we want to study in
this paper, hence the main works done are targeted to the physical
information related to magnetic field of LB and its neighboring
penumbra and umbra. The physical quantities of force-free factor
($\alpha$) and z-component of current ($J_{z}$) and tension force
($T_{z}$), which related to magnetic field and can be deduced from
vector magnetic field observed, are studied specially for this LB.

From the distributions and the amplitudes of $\alpha$, $J_{z}$ and
$T_{z}$ in different parts (namely penumbra, umbra, LB and quiet
region of this active region). it is found that the $\alpha$,
$J_{z}$ and $T_{z}$ of LB are evidently different from those of its
surrounding penumbra and umbra. $\alpha$, $J_{z}$ of LB are larger
than those of its surrounding penumbra and umbra generally. $T_{z}$
of LB are smaller than those of its neighboring umbra, and the
change of sign of $T_{z}$ at the boundary of LB and umbra is very
obvious. These observation results may suggest that magnetic system
of LB and its neighboring umbra (including penumbra) are two
different magnetic systems. The magnetic topologies and environment
of LB and its neighboring umbra are more suitable to create
interactions between these two magnetic systems, such as magnetic
reconnection. For example, the directions of tension force $T_{z}$ of
LB and umbra are opposite, thus, there may exist easily the
instabilities and interactions between these two different magnetic
systems, which may create conditions for the consumptions of magnetic
flux or the redistributions of magnetic field and then result in the
breakup or assembly of sunspots.

\acknowledgments {\it Hinode} is a Japanese mission developed and
launched by ISAS/JAXA, collaborating with NAOJ as a domestic
partner, NASA and STFC (UK) as international partners. Scientific
operation of the {\it Hinode} mission is conducted by the {\it
Hinode} science team organized at ISAS/JAXA. This team mainly
consists of scientists from institutes in the partner countries.
Support for the post-launch operation is provided by JAXA and NAOJ
(Japan), STFC (U.K.), NASA, ESA, and NSC (Norway).
This work was partly supported by the Grants: 2011CB811401, KLCX2-YW-T04, KJCX2-EW-T07, 11373040,
11203036, 11221063, 11178005, 11003025, 11103037, 11103038, 10673016, 10778723 and 11178016, the
Key Laboratory of Solar Activity National Astronomical Observations, Chinese Academy
of Sciences, National Basic Research
Program of China (Grant No. 2011CB8114001) and
the Young Researcher Grant of National Astronomical Observatories,
Chinese Academy of Sciences.


\email{aastex-help@aas.org}.


\begin{thebibliography}{}

\bibitem[Asai(2001)]{asa01} Asai, A., Ishii, T.T., \& Kurokawa, H.
2001, \apj, 555, L65


\bibitem[Berger \& Berdyugina(2003)]{ber03} Berger, T.E. \& Berdyugina, S.V.
2003, \apj, 589, L117

\bibitem[Bharti et al.(2007)]{bha07} Bharti, T., Rimmele, T., Jain, R., Jaaffrey, S.N.A., \&
Smart, R.N. 2007, \mnras, 376, 1291

\bibitem[Bray et al.(1964)]{bra64} Bray, R.J. \& Loughhead, R.E.
1964, Sunspots, The International Astrophysics Series (London:
Chapman \& Hall)

\bibitem[Garcia de La Rosa(1987)]{gar87} Garcia de La Rosa, J.I.
1987, \solphys, 112, 49


\bibitem[Garcia de La Rosa(1987)]{gar87} Garcia de La Rosa, J.I.
1987, \solphys, 112, 49



\bibitem[Jurcak et al.(2006)]{jur06} Jurcak, J., Pillet, V.M., \&
Sobotka, M. 2006, \aap, 112, 49


\bibitem[Katsukawa(2007)]{kat07a} Katsukawa, Y.
2007, New Solar Physics with Solar-B Mission ASP Conference Series,
Edited by Kazunari Shibata, Shin'ichi Nagata, Takashi Sakurai, 369,
p.287

\bibitem[Katsukawa et al.(2007b)]{kat07b} Katsukawa, Y., Yokoyama, T., Berger, T.E., Ichimoto, K.,
Kubo, M., Lites, B.W., Nagata, S., Shimizu, T., A.Shine, R.,
Suematsu, Y., D.Tarbell, T., M.Title, A. \& Sueta, S. Katsukawa, Y.,
2007, \pasj, 59, 577

\bibitem[Kosugi et al.(2007)]{kos07}
 Kosugi, T., Matsuzaki, K., Sakao, T., Shimizu, T., Sone, Y.,
Tachikawa, S., Hashimoto, T., Minesugi, K., Ohnishi, A., Yamada, T.,
Tsuneta, S., Hara, H., Ichimoto, K., Suematsu, Y., Shimojo, M.,
Watanabe, T., Davis, J.M., Hill, L.D., Owens, J.K., Title, A.M.,
Culhane, J.L., Harra, L., Doschek, G.A., \& Golub, L. 2007,
\solphys,  243, 3

\bibitem[Leka(1997)]{lek97} Leka, K.D.
1997, \apj, 484, 900

\bibitem[Liu(2011)]{liu11} Liu, S.
2011, \pasa, online

\bibitem[Muller(1979)]{mul79} Muller, R.
1979, \solphys, 61, 297

\bibitem[Ravindra et al.(2011)]{rav11} 	
Ravindra, B., Venkatakrishnan, P., Tiwari, S.K. \& Bhattacharyya, R.
2011, \apj, 740, 19

\bibitem[Roy(1973)]{roy73} Roy, J.-R.
1973, \solphys, 28, 95

\bibitem[Rudenko \& Anfinogentov(2014)]{rud14} Rudenko, G. V. \& Anfinogentov, S. A.
2014,  \solphys, 289, 1499


\bibitem[Ruedi et al.(1995)]{rue95} Ruedi, I., Solanki, S.K. \&
Livingston, W. 1979, \solphys, 61, 297

\bibitem[Ryutova et al.(2008)]{ryu08}
Ryutova, M., Berger, T., \& Title, A.
2008, \apj, 676, 1356

\bibitem[Schmieder et al.(1994)]{sch94}
Schmieder, B., Hagyard, M. J., Ai, G.X., Zhang, H.Q., Kalman, B., Gyori, L., Rompolt, B., Demoulin, P. \& Machado, M. E.
1994, \solphys, 150, 199

\bibitem[Sobotka et al.(1993)]{sob93} Sobotka, M., Bonet, J.A. \& Vazquez, M.
1993, \apj, 415, 832

\bibitem[Sobotka et al.(1994)]{sob94} Sobotka, M., Bonet, J.A. \& Vazquez, M.
1994, \apj, 426, 404

\bibitem[Shimizu et al.(2009)]{shi09} Shimizu, T., Katsukawa, Y., Kubo, M.,
Lites, B.W., Ichimoto, K., Suematsu, Y., Tsuneta, S., Nagata, S.,
A.Shine, R., \& D.Tarbell, T. 2009, \apj, 696, L66

\bibitem[Tanaka \& Nakagawa(1973)]{tan73} Tanaka, K. \& Nakagawa, Y.
1973, \solphys, 33, 187

\bibitem[Tiwari et al.(2009)]{tiw09} Tiwari, S.T., Venkatakrishnan, P. \& Sankarasubramanian, K.
2009, \apj, 702, L133

\bibitem[Su et al.(2010)]{su10} Su, J.T., Liu, Y., Zhang, H.Q., Mao, X.J., Zhang, Y. \& He, H.
2010, \apj, 170, 710

\bibitem[Lites et al.(1999)]{lit99}
Lites, B. W., Rutten, R. J., Berger, T. E.
1999, \apj, \textbf{517}, 1013.


\bibitem[Spruit \& Scharmer(2006)]{spr06} Spruit, H.C., \& Scharmer, G.B.
2006, \apj, 415, 832


\bibitem[Tsuneta et al.(2008)]{tsu08} Tsuneta, S., Suematsu, Y., Ichimoto, K.,
Shimizu, T., Otsubo, M., Nagata, S., Katsukawa, Y., Title, A.,
Tarbell, T., Shine, R., Rosenberg, B., Hoffmann, C., Jurcevich, B.,
Levay, M., Lites, B., Elmore, D., Matsushita, T., Kawaguchi, N.,
Mikami, I., Shimada, S., Hill, L.,  \& Owens, J. 2008, \solphys,
249,
167


\bibitem[Wang et al.(2008)]{wang08}	
Wang, H.M., Jing, J., Tan, C.Y., Wiegelmann, T. \& Kubo, M.
2008, \apj, 687, 658

\bibitem[Vasquez (1973)]{vas73} Vasquez, M.
1973, \solphys, 31, 377

\bibitem[Venkatakrishnan \& Tiwari(2010)]{ven10} Venkatakrishnan, P. \& Tiwari, S. K.
2010, \aap, 516, L5
\bibitem[Venkatakrishnan et al.(1993)]{ven93} Venkatakrishnan, P., Narayanan, R. S., \& Prasad, N. D. N.
1993, \solphys, 144, 315

\bibitem[Zhang (2010)]{zhang10}	
Zhang, H.Q.
2010, \apj, 716, 1493
\end{thebibliography}
\end{document}